# Modelling real GDP per capita in the USA: cointegration test

I.O. Kitov, O.I. Kitov, S.A. Dolinskaya

**Introduction**

There are several macroeconomic variables crucial for both theoretical consideration and practical usage. Undoubtedly, real economic growth is the most important among them. It defines the speed of economic evolution as associated with the increasing volume and quality of goods and services available for a society as a whole and for every member of the society in particular. Conventional economic theories assume that the growth rate of real GDP reflects routine efforts of every economically active person including those involved in the process of design and control of economic medium. Interactions between economic agents are partly controllable by economic authorities rooting their short-run actions and long-run approaches in the state-of-art economic theory. This theory must describe numerous aspects of the interactions between the regular agents and between the agents and the authorities as well. The literature devoted to various problems of real economic growth is enormous. A comprehensive modern review is available in the "Handbook of Economic Growth" (2005).

There is a simpler and natural explanation using a sole cause for real economic growth. In the framework of the economic concept we have been developing since 2005, the only force driving macroeconomic evolution must be associated with some population group. The intuition behind the concept is inherently related to the observation of the personal income distribution (PID) in the USA. During the years of continuous and relatively accurate measurements between 1960 and 2006, there was no change in the distribution when it is normalized to the total population of 15 years of age and above and nominal per capita GDP growth (Kitov, 2005a, 2005b). This normalization reduces the PID to the portion of the total income obtained by a given portion of the population. The minor change observed in the normalized PID is explained by the change in the age structure of the US society and the increase of the period when age-dependent average income grows with work experience (Kitov, 2005c). Effectively, the PID demonstrates a rigid hierarchy completely reproduced by every new cohort and also by immigrants. The cohort independence is supported by the absence of any significant change with time in the normalized PID in the age groups defined by the US Census Bureau (2002), as found by Kitov (2006a). The rigidity of the PID does not allow any internal part (by age or any other defining property) of population to induce any real acceleration or deceleration of the whole economic system. (Nominal changes are possible, however.) Therefore, a closed economic system can only undergo a constant speed motion. In physics, it is similar to the principle of conservation of linear momentum - particles compiling a closed ensemble and not exposed to external forces can only exchange their momentums but can not change the total momentum of the system. As a consequence, there exists a nonzero economic trend which is defined by a constant annual increment of real GDP per capita, as observed in developed countries (Kitov, 2006c). This constant increment presumes that the growth rate is inversely proportional to the attained level of real GDP per capita. This observation principally differs from a common assumption of economic theories that economic trend is defined by a constant or steady-state growth rate. Such a constant growth rate contradicts the whole bunch of observations in developed countries.

It has been also found that the fluctuations around this constant annual increase are normally distributed (Kitov, 2006c). This indicates that there is no deterministic source of these fluctuations such as changes in demand and supply inspired or/and controlled by some economic authorities. The fluctuations are pure random innovations and can only be related to some exogenous forces. For real economic growth, this force is the change in a single year of age population (Kitov, 2006b). The age is country-specific and is nine years in the USA and UK. In other European countries and Japan the age is eighteen years (Kitov, 2006d).

As a result, one can explicitly formulate a two component model of real economic growth. Empirically, it is based on the observations of the PID in the USA and the normal distribution of annual increments of real GDP per capita in developed countries. This model is absolutely parsimonious since includes only one variable and one constant explaining the whole evolution of real economics as expressed in monetary units. (We do not consider technological evolution as related to the monetary representation.)

So, it is about the time to validate the model by standard statistical and econometric procedures. Juselius and Franchi (2007) (hereto after JF) have proposed an adequate framework for such a validation – the Cointegrated VAR. The principal idea behind the JF's approach is the estimation of statistical properties of the variables defining the models as themselves and in combinations in order to distinguish between possible and unlikely theoretical assumptions. They have also carried out an important initial analysis of conventional theoretical models of economic growth, RBC and DSGE, and found that some principal assumptions underlying the models are not empirically supported. In sense, we follow the JF's procedure and also the procedures we have developed in our previous papers (Kitov, Kitov, & Dolinskaya, 2007a, 2007b).

The JF has established a new standard for the assessment of fundamental economic models. The assumptions behind such models should come and be justified by empirical data not from the easiness of mathematical formulation. At least, the involved variables should meet minimal requirements established by the models themselves. This approach been successfully used in hard sciences and has brought a well-recognized reliability of technical inventions such as aircrafts, bridges, and so on. The reliability follows from an extensive test of each and every parameter, variable, empirical relationship or fundamental law.

We are happy to join the JF's attempt to establish a more rigorous and testable empirical consideration of various economic models and propose as an alternative the two-component model formulated above. Our model describes the data on real GDP per capita in the USA between 1960 and 2002 and allows predictions on economic growth at various time horizons. Accuracy of the predictions depends on the accuracy of population estimates. The model (and corresponding data) is tested in econometric sense and reveals the existence of a (nonlinear) cointegrating relationship between real economic growth and population. The level of confidence associated with the cointegrating relation is high as demonstrated by various statistical tests. The model also involves the lowermost possible number of variables and does not contain any artificial features such as structural breaks. We consider a developed economy as a natural system which evolves according to its own laws – no internal part, including economic authorities, can accelerate the evolution of the system as a whole. Of course, any part of the system can hamper or stop the evolution, as demonstrated by socialist and developing countries.

The remainder of the paper is organized as follows. Section 1 presents the two-component model for real economic growth and the data used in the study. The model is reversed in order to obtain the number of 9-year-olds from the measured economic growth as expressed by real GDP per capita. Section 2 is devoted to the estimation of basic statistical properties of the variables including the order of integration. Section 3 contains three different tests for cointegration between the measured number of 9-year-olds in the USA and that predicted from fluctuation of real economic growth – two associated with the Engle-Granger approach and also the Johansen test. Section 4 presents a number of VAR and VEC models and some estimates of root mean square errors (RMSE) and goodness of fit. Section 5 discusses principal results, also in connection to those obtained by JF, and concludes.

1. **Model and data**

There is a macroeconomic variable characterized by a potential predictability for an evolving economy. This is the annual increment of real GDP per capita (Kitov, 2006b, 2006c). One can distinguish two principal sources of the GDP growth: the change in the number of 9-year-old population and the economic trend associated with per capita GDP values. The trend has the simplest form – no change of the mean annual increment. This is expressed by the following relationship:

$$dG(t)/dt = A \qquad (1)$$

where $G(t)$ is the absolute level of real GDP per capita at time $t$, $A$ is an empirical and country-specific constant. The solution of this ordinary differential equation is as follows:

$$G(t) = At + B \qquad (2)$$

where $B=G(t_0)$, $t_0$ is the starting time of the studied period. Then, the relative growth rate of the real GDP per capita is expressed by the following relationship:

$$g_{trend}(t) = dG/Gdt = A/G \qquad (3)$$

which indicates that the rate is inversely proportional to the attained level of the real GDP per capita and observed growth rate should decay in time with asymptotic value equal to zero.

One principal correction has to be applied to the per capita GDP values published by the Bureau of Economic Analysis (2007). This is the correction for the difference between the total population and the population of 15 years of age and above, as discussed by Kitov (2006b). The concept requires that only this economically active population should be considered when per capita values are calculated.

Following the general concept of the two principal sources of real economic growth (Kitov, 2006b, 2006c) one can write an equation for the growth rate of the real GDP per capita, $g_{pc}(t)$:

$$g_{pc}(t) = 0.5 dN9(t)/(N9(t)dt) + g_{trend}(t) \qquad (4)$$

where $N9(t)$ is the number of 9-year-olds at time $t$. One can obtain a reversed relationship defining the evolution of the 9-year-old population as a function of economic growth:

$$d(lnN9(t)) = 2(g_{pc} - A/G(t))dt \qquad (5)$$

Relationship (5) defines the evolution of the 9-year-olds as described by the GDP per capita growth rate. The start point of the evolution has to be characterized by some initial value of population. Various population estimates (for example, post- and intercensal one) potentially require different initial values and coefficient $A$. Instead of integrating (5) analytically, we use a discrete form for the annual readings of all the involved variables, which can be rewritten as follows:

$$N9(t+\Delta t) = N9(t)[1 + 2\Delta t(g_{pc}(t) - A/G(t))] \qquad (6)$$

where $\Delta t$ is the time unit equal to one year. Relationship (6) uses is a simple representation of the time derivative of the population estimate, where the derivative is approximated by its estimate at point $t$. The time series $g_{pc}$ and $G$ are independently measured variables. In order to obtain the best prediction of the $N9(t)$ one has to vary coefficient $A$ and the initial value - $N9(t_0)$. The best-fit parameters can be obtained by some standard technique minimising the RMS difference between the predicted and measured series. In this study, only visual fit between the curves is used, with the average difference minimised to zero. This approach might not provide the lowermost standard deviation, however.

Relationship (6) can be interpreted in the following way - the deviation between the observed growth rate of GDP per capita and that defined by the long-tern trend is completely defined by the change rate of the number of 9-year-olds. A reversed statement is hardly to be correct - the number of people of some specific age can not be completely or even in large part defined by contemporary real economic growth. For example, the causality principle prohibits the present to influence the birth rate nine years ago. Econometrically speaking, the number of 9-year-olds has to be a weakly exogenous variable relative to current economic growth. This property of the variables is used in the VAR models in Section 4.

In fact, (6) provides a prediction for the number of 9-year-olds using only independent measurements of real GDP per capita. Therefore, amplitude and statistical properties of the deviation between the measured and predicted number of 9-year-olds can serve for validation of (4) and (5). Therefore, in Sections 2 through 4 we use the predicted number of 9-year-olds for statistical estimates instead of the real GDP per capita readings themselves. The link between population and economic growth is effectively nonlinear and would be difficult to study in the assumption of a linear relation. Because both involved variables are measured with some uncertainty and probably are nonstationary the Cointegrated VAR analysis, proposed by JF, should be appropriate.

There are numerous revisions and vintages of population estimates. Figure 1 compares post- and intercensal population estimates of the number of 9-year-olds between 1960 and 2002 (US CB, 2007). The error of closure for the 9-year-olds, i.e. the difference between the census count and the postcensal estimate at April 1, 2000, is 57233. (The error of closure for the population group between 5 and 13 years of age is 1309404, however, i.e. approximately twice as large for every single year of age as that for the 9-year-olds.) The difference is proportionally distributed over the 3653 days between April 1, 1990 and April 1, 2000 (US CB, 2004). Therefore, the level of the intercensal estimate is represented by the level of the postcensal one plus corresponding portion of the error of closure. The curves in Figure 1 demonstrate a growing divergence between the estimates. (There are also some minor corrections between adjacent years of birth in wider age groups.) After April 2000, both estimates in Figure 1 are apparently postcensal with different bases in 2000. Even this minor deviation between the estimates might be of importance for statistical tests and inferences and both are analyzed in this study.

The real GDP per capita readings are obtained using the total real GDP and the number of people of 15 years of age and above. This excludes from the macroeconomic consideration those who do not add to the economic growth (Kitov, 2006c). Figure 2 depicts the growth rate of the real GDP per capita in the USA between 1960 and 2002 used in the study. The average growth rate is 0.0204 with standard deviation of 0.0217. There are seven negative readings coinciding with the recession periods defined by the National Bureau of Economic Research (NBER, 2007). There was no negative growth in 1960, however.

The period between 1960 and 2002 has been chosen by the following reasons. Before 1960, the single year of age population estimates are not reliable and might introduce a significant distortion in statistical estimates and inferences. After 2002, the GDP values are prone to comprehensive NIPA revision of unknown amplitude, which historically occurred about every 5 years (Fixler & Green, 2005). The most recent comprehensive revision was in 2003 and spanned the years between 1929 and 2002.

## 2. Unit root tests

The technique of linear regression for obtaining statistical estimates and inferences related to time series is applicable only to stationary series, as Granger and Newbold (1967) showed forty years ago. Two or more nonstationary series can be regresses only in the case when there exists a cointegrating relation between them (Hendry & Juselius, 2001). Therefore, the first step in the econometric studies of time dependent data sets is currently consists is estimation of the order of integration of the involved series. Unit root tests applied to the original series and their first and higher order differences are a useful tool to determine the order of integration.

Standard econometric package Stata9 provides a number of appropriate procedures implemented in the interactive form. The Augmented Dickey-Fuller (ADF) and the modified DF t-test using a generalized least-squares regression (DF-GLS) are used in this study. Potentially, the tests provide the most adequate results for the available short series consisting of only 41 annual readings - the real GDP per capita and the number of 9-year-olds. In any case, small samples usually characterized by a limited reliability of statistical inferences.

There are four original time series tested for unit roots - the measured and predicted according to (6) number of 9-year-olds between 1962 and 2002. Each of the series contains two versions - a postcensal and intercensal one for the period between 1990 and 2000, i.e. between two decennial censuses. The difference is minor, as Figure 1 demonstrates, but the intercensal series potentially contains such artificial features as autocorrelation introduced by the Census Bureaus during the revision associated with the error of closure. Therefore, the postcensal time series potentially is statistically "better" than the intercensal one.

Some results of unit root tests for the four original series are listed in Table 1. All these series are characterized by the presence of unit roots - the test values are significantly larger than the 5% critical values. In the ADF tests, trend specification is constant and the maximum lag order is 3. In the DF-GLS tests, the maximum lag is 4 and the same trend specification is used. Hence, one can conclude that the studied time series are nonstationary. The order of integration is not clear, however.

The first differences of the measured and predicted number of the 9-year-olds (postcensal version) between 1962 and 2002, i.e. the reading for 1961 is also used, are presented in Figure 3. There is no significant trend in the data and one can presume a constant as trend specification. The average value is 9600 and 12625 and standard deviation is 152487 and 105287 for the measured and predicted time series, respectively.

Table 2 summarizes the results of corresponding unit root tests. The predicted time series are definitely characterized by the absence of unit roots, as the ADF and DF-GLS both demonstrate for the maximum lag order 2. For lag 3, the ADF gives values just marginally below the 5% critical value. The measured time series have specific autoregressive properties intrinsically related to the methodology of population revisions and are characterized by mixed results of the unit root tests. The DF-GLS test rejects the null hypothesis of the presence of a unit root for all lags from 1 to 3. The ADF rejects the null only for lag 0. Bearing in mind the shape of the measured original curves in Figure 1, which demonstrate a quasi-sinusoidal behaviour without any significant linear trend, one can assume that their first differences are stationary. In this study, the absence of unit roots in all the first difference series is accepted.

The presence of the unit roots in the original series and the absence of unit roots in the first differences evidences that the former series are integrated of order 1. This fact implies that a cointegration analysis has to be carried out before a linear regression because the latter is potentially a spurious one.

3. **Cointegration test**

The assumption that the measured number of 9-year-olds in the USA and that predicted from the real economic growth, as expressed by the growth rate of per capita GDP, are two cointegrated non-stationary time series is equivalent to the assumption that their difference, $\varepsilon(t) = N9_m(t) - N9_p(t)$, is a stationary or I(0) process. The predicted and measured series are shown in Figures 4 and 6, and their differences in Figures 5 and 7.

It is natural to start with unit root tests in the difference. If $\varepsilon(t)$ is a non-stationary variable having a unit root, the null of the existence of a cointegrating relation can be rejected. Such a test is associated with the Engle-Granger approach (1987), which requires $N9_m(t)$ to be regressed on $N9_p(t)$ as the first step, however. It is worth noting, however, that the predicted variable is obtained by a procedure similar to that of linear regression and provides the best visual fit between corresponding curves. The Engle-Granger approach is most reliable and effective when one of the two involved variables is weakly exogenous, i.e. is driven by some forces not associated with the second variable. This is the case for the per capita GDP and the number of 9-year-olds. The latter variable is hardly to be driven by the former one. The existence of an opposite causality direction is the aim of this study.

The results of the ADF and DF-GLS tests listed in Table 3 indicate the absence of a unit root in the measured-predicted difference series for both post- and intercensal population estimates. Since the predicted series are constructed in the assumption of a zero average difference, trend specification in the tests is *none*. The maximum lag order in the tests is 3. The test results give strong evidences in favor of the existence of a cointegrating relation between the measured and predicted time series. Thus, from econometric point of view it is difficult to deny that the number of 9-year-olds is potentially the only defining factor behind the observed fluctuations of the real economic growth around the economic trend associated with constant annual increment of GDP per capita.

The next step is to use the Engle-Granger approach again and to study statistical properties of the residuals obtained from linear regressions of the measured and predicted single year of age populations. A pitfall of the regression analysis consists in the time shift between the measured and predicted series – the former corresponds to July 1 and the latter to December 31 of the same year. This phase shift results in deterioration of regression results but can not be recovered since only annual population estimates are available before 1980. Table 4 summarizes principal results of relevant unit root tests with the same specifications as accepted for the difference of the same series. The null hypothesis of a unit root presence is rejected for both time series and all lags. Thus, the residuals build an I(0) time series - the predicted and measured variables are cointegrated.

The Johansen (1988) approach is based on the maximum likelihood estimation procedure and tests for the number of cointegrating relations in the vector-autoregressive representation. The Johansen approach allows simultaneous testing for the existence of cointegrating relations and determining their number (rank). For two variables, only one cointegrating relation is possible. When cointegration rank is 0, any linear combination of the two variables is a non-stationary process. When rank is 2, both variables have to be stationary. When the Johansen test results in rank 1, a cointegrating relation between the involved variables does exist.

In the Johansen approach, on has first to analyze specific properties of the underlying VAR model for the two variables. Table 5 lists selection statistics for the pre-estimated maximum lag order in the VAR. Standard trace statistics is extended by several information criteria: the final prediction error, FPE, the Akaike information criterion, AIC, the Schwarz Bayesian information criterion -SBIC, and the Hannan and Quinn information criterion, HQIC. All tests and information criteria indicate the maximum pre-estimated lag order 1 for VARs and VECMs. Therefore, maximum lag order 1 was used in the Johansen tests along with constant as trend specification.

Properties of the VAR error term have a critical importance for the Johansen test (Hendry & Juselius, 2001). A series of diagnostic tests was carried out for the VAR residuals. The Lagrange multiplier test for the postcensal time series resulted in $\chi^2$ of 0.34 and 0.09 for lags 1 and 2 respectively. The test accepts the null hypothesis of the absence of any autocorrelation at given lags. The Jarque-Bera test gives $\chi^2$=7.06 (Prob>0.03) with skewness=0.96 and kurtosis=3.77, skewness being of the highest importance for the Jarque-Bera normality test and the validity of statistical inference. Hence, the residuals are probably not normally distributed, as expected from the artificial features of the measured population time series. The VAR model stability is guaranteed by the eigenvalues of the companion matrix lower than 0.63. As a whole, the VAR model accurately describes the data and satisfies principal statistical requirements applied to the residuals.

Table 6 represents results of the Johansen tests – in both cases cointegrating rank is 1, i.e. there exists a long-run equilibrium relations between the measured and predicted number of 9-year-olds in the USA. The predicted number is obtained solely from the level of real GDP per capita. (The growth rate is just the time derivative normalized to the level.) We do not test for causality direction between the variables because the only possible way of influence, if it exists, is absolutely obvious.

In this Section, three different tests have demonstrated at a relatively high level of confidence that the measured and predicted number of 9-year-olds in the USA are cointegrated. In addition to the existence of a long-run relation between the single year of age population and real economic growth, one can use the relation as a reliable fundament for the prediction of the economic evolution in the USA.

## 4. The VAR, VECM, and linear regression

Now we are sure that the measured and predicted single year of age population series are cointegrated. Therefore, the estimates of the goodness-of-fit and RMSE in various statistical representations have to be valid and provide important information on the accuracy of corresponding measurements and the relation itself.

A VAR representation provides a good estimate of $R^2$ and RMSE due to strong noise suppression. In practice, AR is a version of a weighted moving average, which optimizes noise suppression throughout the whole series. Two VAR models are possible, however, with the predicted time series used as an exogenous predictor and as an endogenous variable. Table 7 summarizes results of the VAR models and demonstrates that the goodness of fit is excellent (and reliable), with the highest $R^2$~0.95 and the lowermost RMSE near 72000 corresponding to the exogenous predicted time series for the postcensal population estimates. This version of VAR uses the maximum lag order 2, and the Table confirms that coefficient L2 is not significant in line with the previous estimates of the maximum lag. The coefficient for the predictor is significant.

A VECM representation uses additional information to that provided by the VAR models due to separation of noise and equilibrium relationship. So, it potentially provides an improvement on the VAR models. Table 8 lists some results obtained in the VECM (Cointegrated VAR) representation. Coefficient β defining the link between the measured and predicted series is significant in both cases confirming the existence of a cointegrating relation. Coefficients $α_1$ and $α_2$ define the input of the cointegrating relation to the I(0) time series of lagged first differences of the measured and predicted series. Their estimates are significant and show a relatively large error correction effect. Coefficients of the LD terms are both insignificant as corresponded to the largest lag order 2. The values of $R^2$ are relatively high (0.34 and 0.32) and RMSE is ~90000 and 130000 for the postcensal and intercensal series, respectively. The RMSE values are slightly larger than those produced by the VAR models.

Finally, Table 9 presents some results obtained by linear regression. The results are biased by the time shift between the series and are inferior to those obtained using VAR and VECM. Moving average technique, however, provides a slight improvement in statistical estimates. This effect is inherently related to noise suppression in the time series.

Despite a very high goodness of fit approaching 0.95 in the VAR representation, the RMSE estimates are relatively large. This severely complicates the usage of (4) for the prediction of real economic growth in the USA. The RMSEs, however, are comparable in amplitude with the uncertainty of the population estimates, especially at younger ages (West & Robinson, 1999). In addition, a conservative estimate of the GDP measurement error is of 0.5 % to 1%, which includes also the uncertainty associated with CPI and GDP deflator. In order to distinguish between the measurement errors and some true deviations from the cointegrating relation one needs a substantial improvement in population estimates. This is a standard situation in hard sciences, and now might be in economics as well if the JF initiative will be adequately supported.

## 5. Discussion and conclusions

There is an equilibrium (nonlinear) long-run relationship between the number of 9-year-olds and the real per capita GDP. This fact implies that real economic growth, as expressed in monetary units, is driven only by the age structure of the US society. An increasing number of 9-year-olds would guarantee an accelerating growth extra to that defined by the constant annual increment of the real GDP per capita, and vice versa.

The period of the oscillations in the number of 9-year-olds in the USA is approximately 30 years between the peaks in 1970 and 2000. Such long-period oscillations in economic evolution are well-know since the 1920s, when Russian economist Kondratiev published his original analysis. Our model gives a natural explanation of the Kondratiev waves – they are related to the natural increases and decreases in birth rate (and migration). For obvious reasons, the birth rate can not increase or decrease monotonically – it would result in exponential population growth or in a complete extinction of the population. Hence, the cycles are observed.

A bad news for the USA is that the next fifteen years will be probably associated with the decreasing branch of the K-wave. Taking into account the effect of the decreasing background growth rate associated with the increasing real GDP per capita in (3), one can expect a significant deceleration in the US economy as expressed by a lower growth rate of real GDP per capita.

However, if the total population continues to grow at an annual rate of 1 per cent, as has been observed in the USA during the last forty years, the negative effect of the relative deceleration will be reduced. In developed European countries, the effect of the total population growth is practically negligible and they seemingly do not grow so fast as the USA does. This is just confusion if to use per capita GDP values.

The fluctuations of the annual increment of real GDP per capita around a constant level represent a random process. This stochastic component is driven only by one force and can be actually predicted to the extent one can predict the number of 9-year-olds at various time horizons. The population estimates for younger ages in previous years provide an excellent source for the prediction of the number of 9-year-olds. The growth rate of a single year population can be predicted with even higher accuracy because the levels of adjacent cohorts change practically in sync. Therefore, the number of 7-year-olds today is a very good approximation of the number of 9-year-olds in two years. Theoretically, one can use the populations for an accurate prediction. In practice, modern methodology of the population estimates does not provide an adequate accuracy and only long-term changes have high enough amplitude for a reliable resolution of the link between real economic growth and population, as Figures 4 and 6 illustrate.

The concept we are developing links the fluctuations of growth rate to the people outside the current economic structure. They bring to the system a changing and nonzero input, which can be interpreted as new demand for goods and services. Those economic agents who are currently inside the system can not change real demand per capita due to the rigid PID. Immigrants and the population decrease associated with deaths also can not change per capita GDP values because the PID does not demonstrate any effect of these potential sources of changes. One can presume that the hierarchy of personal incomes momentarily recovers to its origin structure when accommodating the disturbances induced by these two sources.

Our concept is strongly supported by the results in JF. They have found that the principal source of economic variations is the demand for consumption and for labor but not shocks to technology or total factor productivity. Newcomers, as represented by 9-year-olds, somehow introduce their long-term demand for consumption into the economic system. This demand changes with time according to the changes in the number of 9-year-olds and induces relevant changes in the demand for labor. A complication to conventional models is the decaying economic trend, as defined by (3).

Actual developed economies do not consist of two distinct parts, which are usually described as investment and consumption, the former being the driving force of shocks to technology and total factor productivity. Theories of endogenous economic growth, however, are based on this assumption and stress the importance of investment for the rate of economic growth. In our framework, there is no link between the size of an economy, as expressed in monetary units, and its technological content. In other words, any set of technological breakthroughs achieved during a certain period, for example one year, has the same money valuation. What important for the monetary size is only changes in quantitative characteristics of population – the age structure. We also do not share the opinion or assumption that investments are made for the sake of economic growth as itself. One hardly can imagine that an owner, shear holder or manager who really wants an overall economic growth and decides what input s/he can bring to the process. Investment decisions are rather made on for a sole purpose, which is psychologically economically justified, one wishes by all means to elevate the current position in PID. Technological innovations, in broad sense not only purely technological but also cultural increasing diversity of services, allow getting a tool to depose some people from their top positions in the PID, which is rigid as we have found before. The rigidity does not permit to join a top position – only exchange of positions is possible. This is a mobile process with variations in outcome. By the way, technologically excellent discoveries do not always guarantee income increase – another important feature is the capability to convince people to pay for the products.

So, the only purpose to invest is to progress in the income pyramid to higher steps. This is a routine, strong and long-run interest and demand. Sometimes it uses not the best sides of human psychology and reflexes. But, in general, it makes what it has to make – brings random and deterministic innovations in technologies. Juselius and Franchi (2007) justified our concept by empirical analysis. No technological innovations induce fluctuations in economic growth as expressed in monetary units. (We do not consider here technical policy aimed to select the soundest innovations, which can definitely bring a better result for the society as a whole. For example, investments in military technologies brought a large-scale profit to many areas of civil techniques.) JF denies the possibility of technology, whatever it is, to drive economic part of social life.

The Great Moderation is easily explained in our framework. Amplitude of the fluctuations of the defining age population around the constant level has been decaying since the 1980s, as Figure 5 and 7 demonstrate. (Reasons behind the smoothing of the population changes are beyond the scope of this study but deserve a special attention.) The GDP growth rate (with the increasing per capita GDP level as denominator) suffers an additional decrease. Because inflation in the USA is driven by the change in the level of labor force (Kitov, 2006e, 2006f, 2007; Kitov, Kitov, & Dolinskaya, 2007a, 2007b) the decreasing demand for labor associated with the decreasing demand for consumption leads to lower inflation. The Great Moderation is not going to leave the scene in the future.

**Tables**

Table 1. Unit root tests for the measured and predicted number of 9-year-olds. Trend specification is constant.

| Test | Lag | intercensal | | postcensal | | 1% critical |
|---|---|---|---|---|---|---|
| | | predicted | measured | predicted | measured | |
| ADF | 0 | -1.50 | -0.72 | -1.51 | -0.70 | -3.65 |
| | 1 | -2.10 | -1.39 | -2.10 | -1.40 | -3.66 |
| DF-GLS | 1 | -2.34 | -1.52 | -2.35 | -1.55 | -2.63 |
| | 2 | -1.82 | -1.60 | -1.82 | -1.62 | -2.63 |

Table 2. Unit root tests for the first differences of the measured and predicted number of 9-year-olds. Trend specification is constant. The maximum lag order is 3.

| Test | Lag | Postcensal | | Intercensal | | 1% critical |
|---|---|---|---|---|---|---|
| | | predicted | measured | predicted | measured | |
| ADF | 0 | -4.86* | -4.22* | -4.87* | -4.27* | -3.65 |
| | 1 | -4.66* | -3.37 | -4.67* | -3.39 | -3.66 |
| | 2 | -3.86* | -2.80 | -3.86* | -2.80 | -3.66 |
| | 3 | -3.44 | -3.22 | -3.44 | -3.20 | -3.67 |
| DF-GLS | 1 | -4.64* | -3.01* | -4.54* | -3.02* | -2.63 |
| | 2 | -3.67* | -2.48 | -3.67* | -2.48 | -2.63 |
| | 3 | -3.12* | -2.84* | -3.12* | -2.84* | -2.63 |

Table 3. Unit root tests for the differences between the measured and predicted number of 9-year-olds. Trend specification is constant. The maximum lag order is 3.

| Test | Lag | Time series | | 1% critical |
|---|---|---|---|---|
| | | postcensal | intercensal | |
| ADF | 0 | -2.87* | -2.85* | -2.64 |
| | 1 | -3.67* | -3.59* | -2.64 |
| | 2 | -2.99* | -3.92* | -2.64 |
| | 3 | -2.90* | -2.83* | -2.64 |
| DF-GLS | 1 | -3.55* | -3.47* | -2.64 |
| | 2 | -2.98* | -2.92* | -2.64 |
| | 3 | -2.92* | -2.85* | -2.64 |

Table 4. Unit root tests for the residual time series of a linear regression of the measured series on the predicted one. The measured and predicted series are the numbers of 9-year-olds. Trend specification is none (zero average value of the residuals) and maximum lag order 3.

| Test | Lag | Time series | | 1% critical |
|---|---|---|---|---|
| | | postcensal | intercensal | |
| ADF | 0 | -3.03* | -3.02* | -2.64 |
| | 1 | -3.88* | -3.86* | -2.64 |
| | 2 | -3.15* | -3.13* | -2.64 |
| | 3 | -3.05* | -3.01* | -2.64 |
| DF-GLS | 1 | -3.71* | -3.69* | -2.64 |
| | 2 | -3.06* | -3.04* | -2.64 |
| | 3 | -2.98* | -2.95* | -2.64 |

Table 5. Pre-estimation lag order selection statistics. All tests and information criteria indicate the maximum lag order 1 as an optimal one for VARs and VECMs.

|  | Lag | LR | FPE | AIC | HQIC | SBIC |
|---|---|---|---|---|---|---|
| postcensal | 1 | 63.03* | 5.8e+09* | 25.31* | 25.36* | 25.44* |
| intercensal | 1 | 61.63* | 6.1e+09* | 25.38* | 25.42* | 25.51* |

FPE    - the final prediction error,
AIC    - the Akaike information criterion,
SBIC   - the Schwarz Bayesian information criterion,
HQIC  - the Hannan and Quinn information criterion

Table 6. Johansen test for cointegration rank for the measure and predicted time series. Trend specification is constant. Maximum lag order is 2.

| Time series | Rank | Eigenvalue | SBIC | HQIC | Trace statistics | 5% critical value |
|---|---|---|---|---|---|---|
| postcensal | 1 | 0.39747 | 52.4803* | 52.23414* | 2.1984* | 3.76 |
| intercensal | 1 | 0.37905 | 52.55018* | 52.30402* | 2.1177* | 3.76 |

Table 7. VAR models for the measured and predicted number of 9-year-olds for the postcensal and intercensal estimates. Maximum lag order is 2. Two cases for the predicted time series are considered - endogenous and exogenous one.

| Measured-Predicted VAR | RMSE | $R^2$ | Measured | | Predicted | | |
|---|---|---|---|---|---|---|---|
| | | | L1 | L2 | L0 | L1 | L2 |
| exogenous - postcensal | 71645 | 0.9489 | 0.82* [0.13] | -0.12 [0.12] | 0.34* [0.06] | - | - |
| endogenous -postcensal | 89300 | 0.9229 | 0.85* [0.19] | -0.17 [0.15] | - | 0.33* [0.11] | -0.03 [0.12] |
| exogenous - intercensal | 73954 | 0.9474 | 0.82* [0.13] | -0.11 [0.12] | 0.35* [0.06] | - | - |
| endogenous - intercensal | 92440 | 0.9202 | 0.88* [0.19] | -0.17 [0.16] | - | 0.33* [0.11] | -0.05 [0.12] |

Table 8. VECM for the postcensal and intercensal estimates of the number of 9-year-olds. The maximum lag is 2. Cointegrating rank 1 for the relationship between the measured and predicted time series.

| Measured-Predicted VECM | RMSE | $R^2$ | $\beta$ | $\alpha_1$ | $\alpha_2$ | Measured LD | Predicted LD |
|---|---|---|---|---|---|---|---|
| postcensal | 89839 | 0.3446 | -1.21* [0.11] | -0.24* [0.10] | 0.28 [0.17] | 0.11 [0.16] | 0.06 [0.13] |
| intercensal | 93007 | 0.3181 | -1.24* [0.12] | -0.22* [0.10] | 0.29 [0.16] | 0.11 [0.16] | 0.08 [0.13] |

Table 9. Results of linear regression of the measured time series on the predicted one.

| Time series | Regression | Tangent | Constant | $R^2$ | RMSFE |
|---|---|---|---|---|---|
| postcensal | M vs. P | 0.85* [0.09] | 569325 [326128] | 0.71 | 160000 |
|  | M vs. MA(2) | 0.94* [0.07] | 221197 [274652] | 0.81 | 130000 |
|  | M vs. MA(3) | 1.09* [0.06] | -318464 [231765] | 0.89 | 99985 |
| intercensal | M vs. P | 0.86* [0.09] | 511114 [330855] | 0.72 | 160000 |
|  | M vs. MA(2) | 0.96* [0.06] | 167488 [281940] | 0.81 | 130000 |
|  | M vs. MA(3) | 1.04* [0.07] | -126233 [249095] | 0.86 | 110000 |

M - measured time series
P - predicted time series
MA(.) – moving average

**Figures**

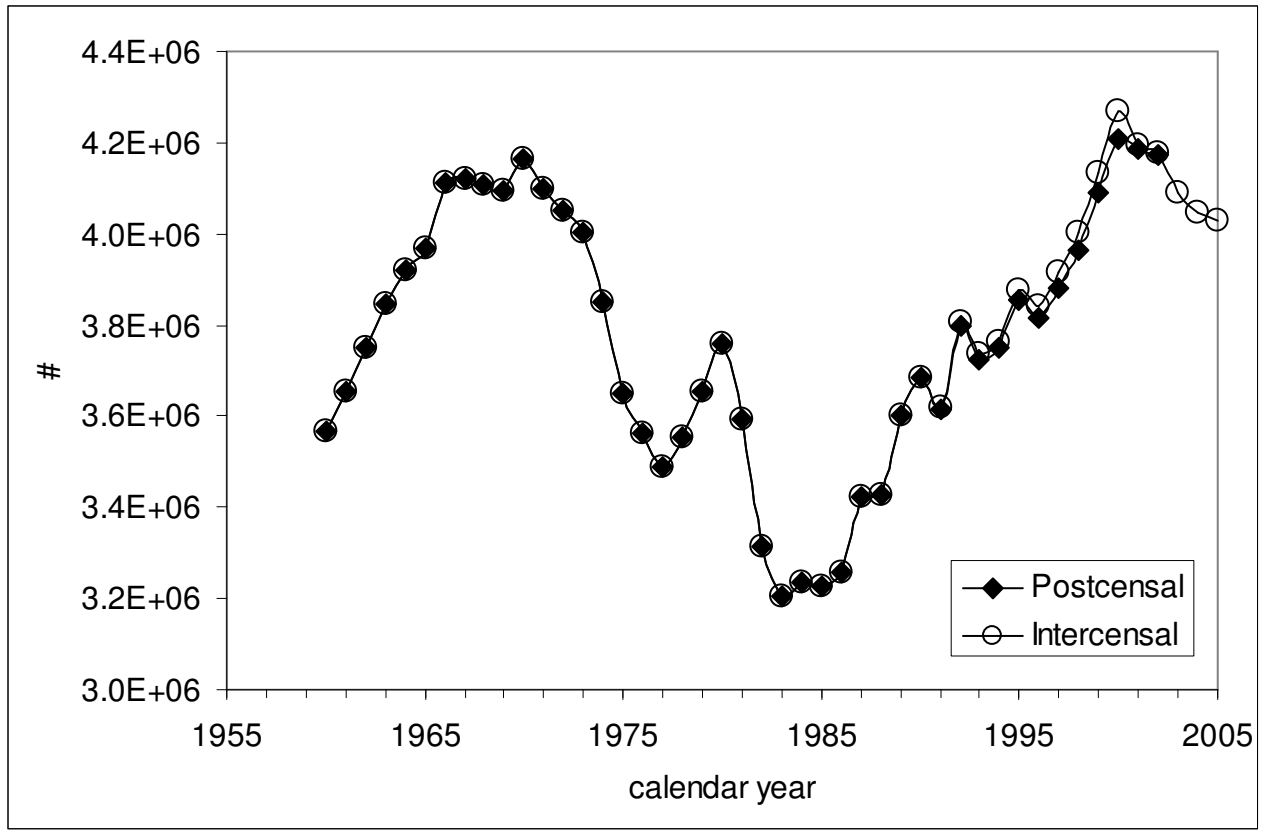

Figure 1. Comparison of the postcensal and intercensal estimates of the number of 9-year-olds reported by the US Census Bureau (2007). The difference is observed only during the years between 1990 and 2002.

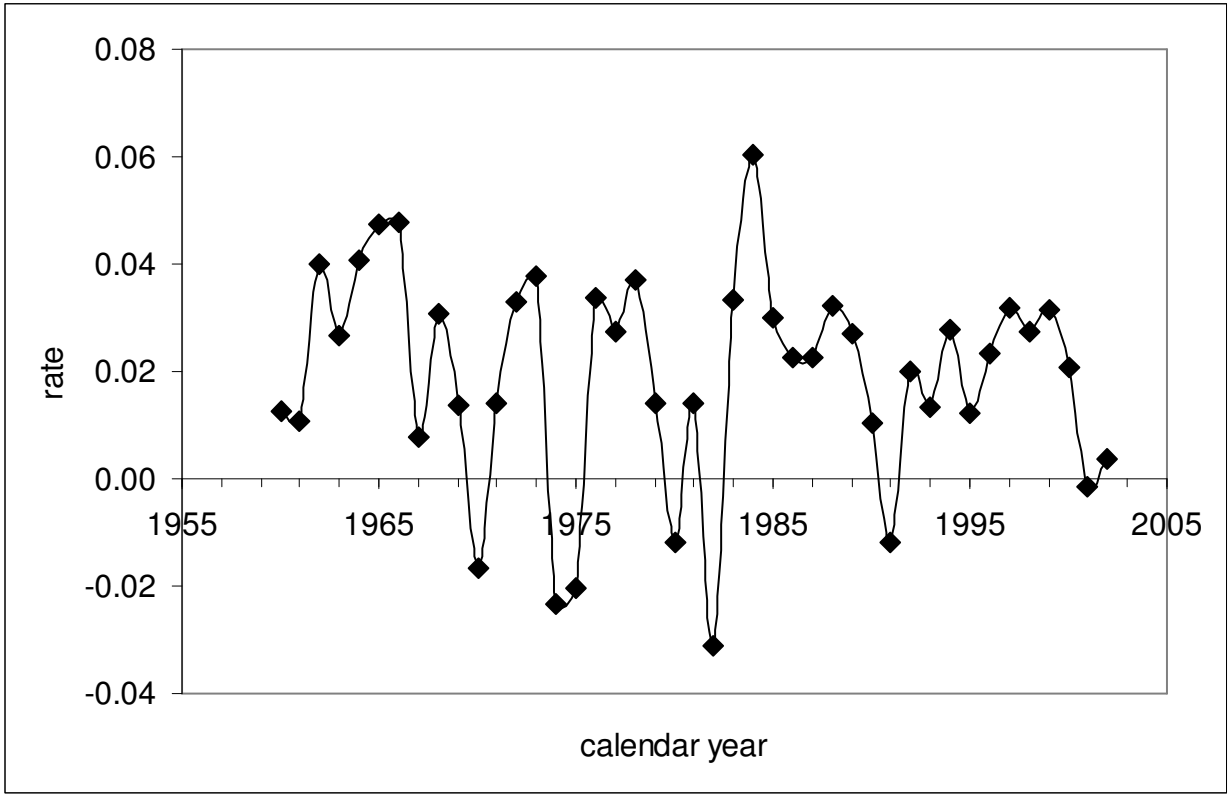

Figure 2. The growth rate of the real GDP per capita in the USA between 1960 and 2002. The growth rate is corrected for the difference between total population and that above 15 years of age (Kitov, 2006b). The average growth rate is 0.0204 with standard deviation of 0.0217. There are seven years of negative growth.

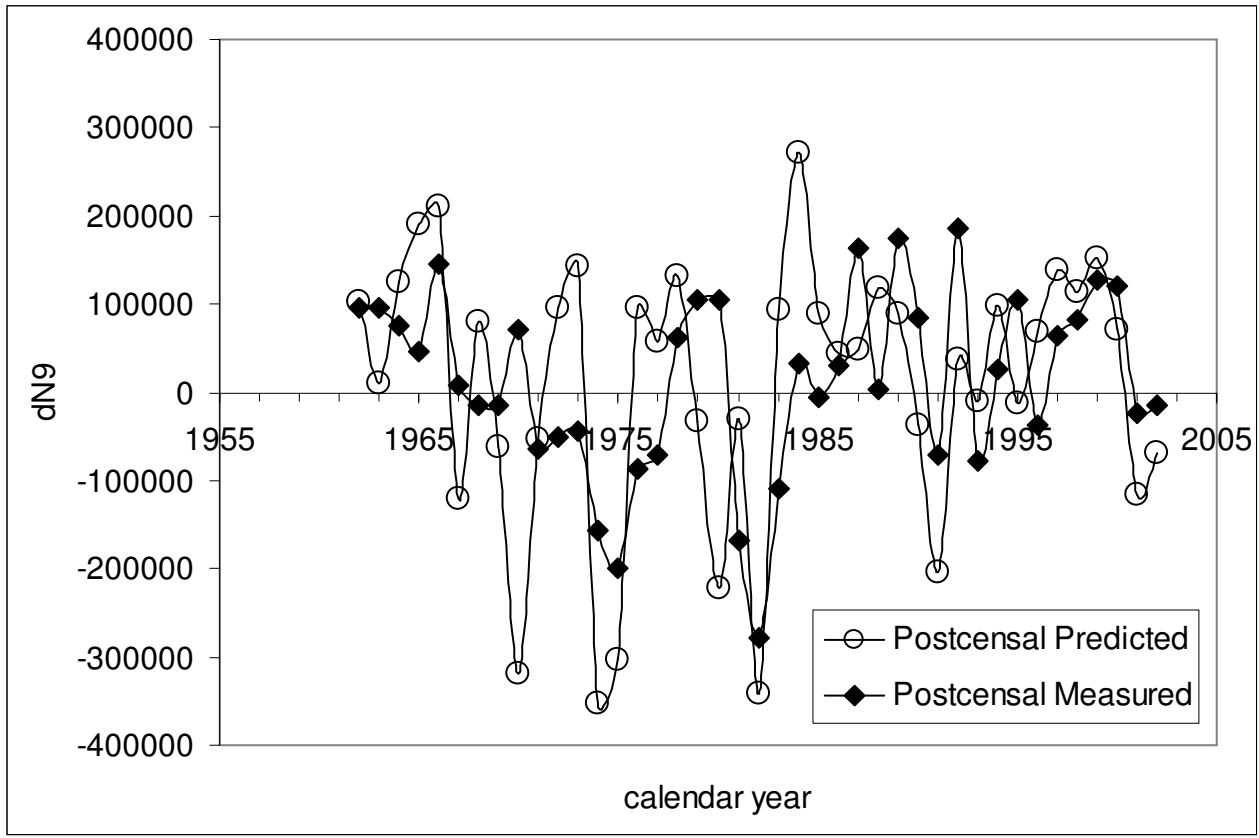

Figure 3. The first differences of the measured (postcensal) and predicted number of 9-year-olds. There is not significant trend in the time series with average values 9600 and 12625, respectively. (Standard deviation is 152487 and 105287.)

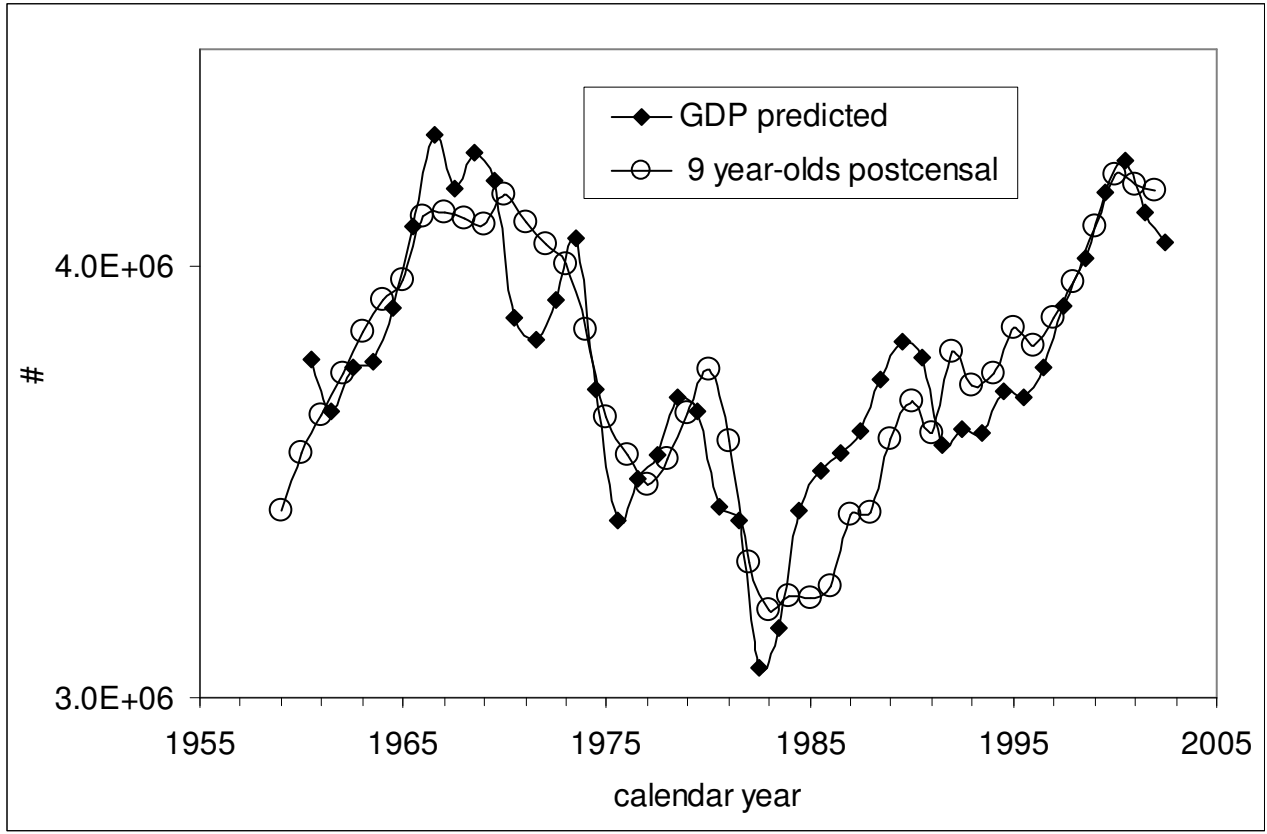

Figure 4. Comparison of the measured and predicted postcensal population estimates between 1960 and 2002.

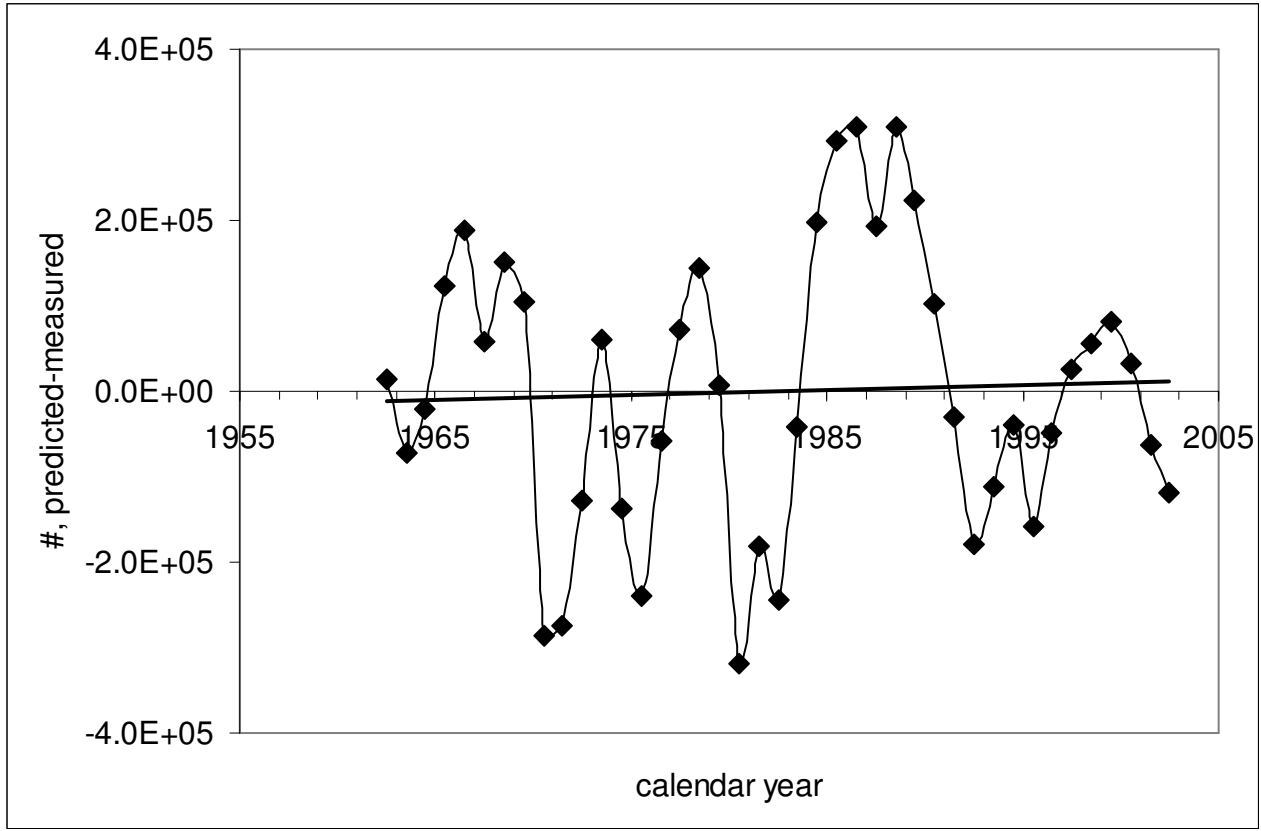

Figure 5. The difference between the measured and predicted population estimates presented in Figure 4. For the period between 1962 and 2002, the average difference is 0 and standard deviation is 164926 for coefficient $A$=547.1325 and the initial value for the population of 3900000 in 1959. Linear regression is represented by a bold straight line.

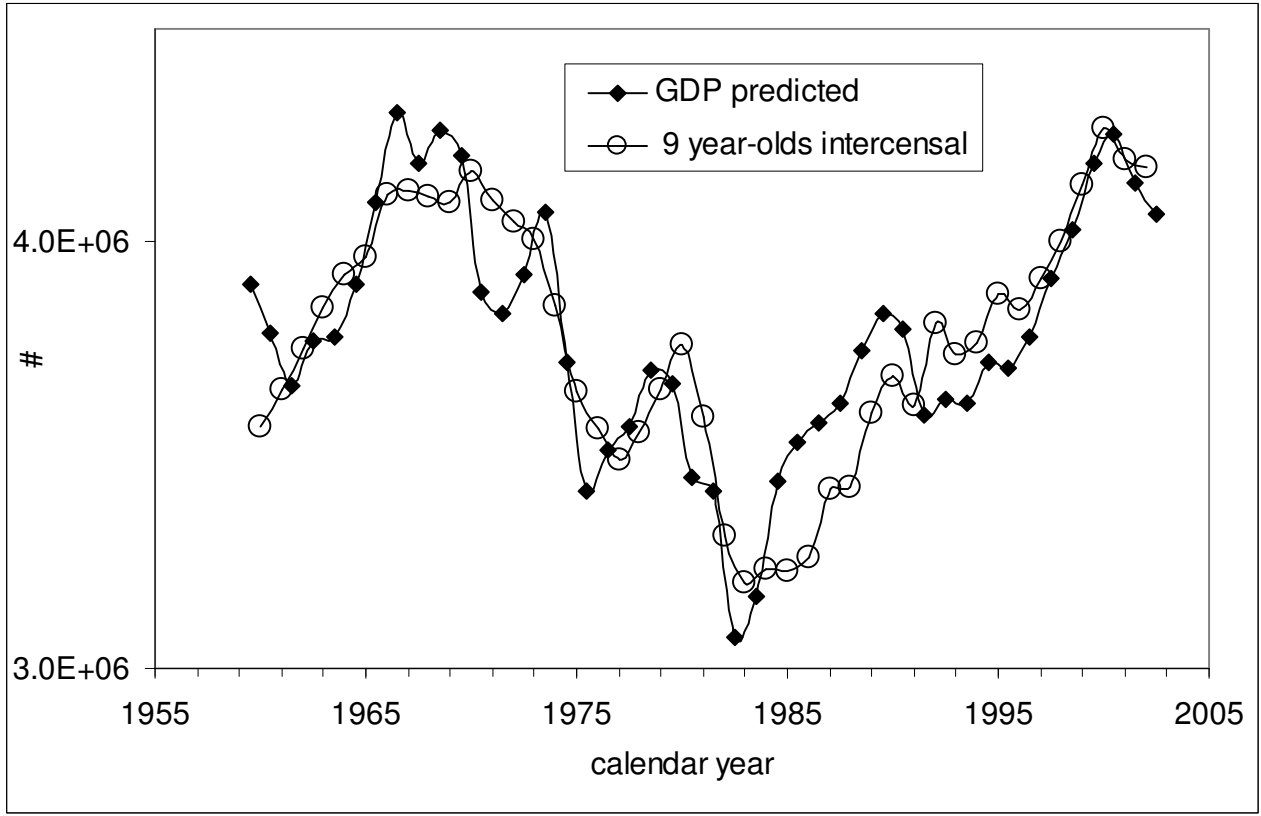

Figure 6. Comparison of the measured and predicted intercensal population estimates between 1960 and 2002.

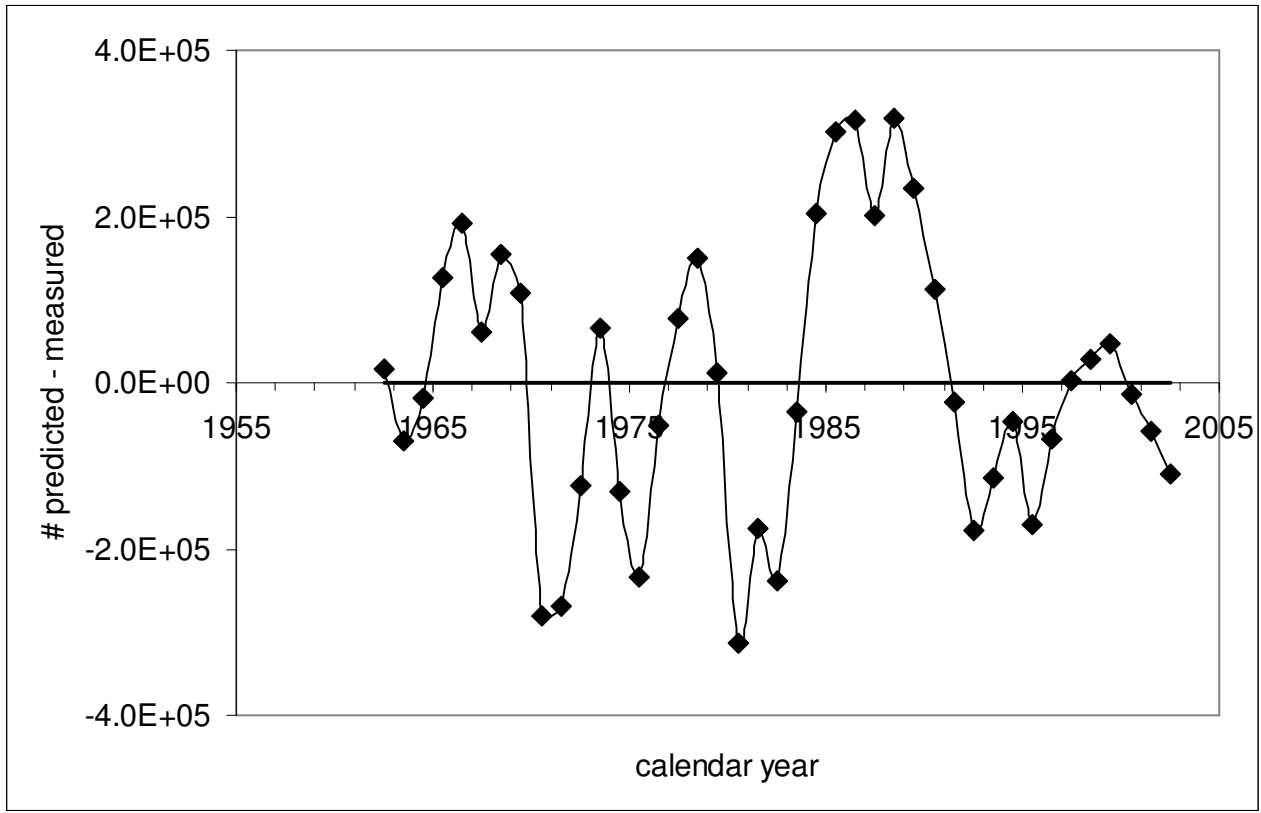

Figure 7. The difference between the measured and predicted population estimates presented in Figure 6. For the period between 1962 and 2002, the average difference is -1 and standard deviation is 165744 for coefficient *A*=546.079 and the initial value of population of 3900000 in 1959. Linear regression is represented by a bold straight line.